\documentclass[aps,prl,12pt,preprint]{revtex4}

\usepackage{graphicx}
\usepackage{dcolumn}
\usepackage{bm}
\usepackage{amsmath}
\usepackage{amssymb}
\usepackage{footnpag}

\begin{document}
\title{Kondo Behavior of U in CaB$_6$}

\author{G.A. Wigger}
\author{E. Felder}
\author{S. Weyeneth}
\author{H. R. Ott}

\affiliation{Laboratorium f\"{u}r Festk\"{o}rperphysik,
ETH-H\"{o}nggerberg, CH-8093 Z\"{u}rich, Switzerland }

\author{Z. Fisk}

\affiliation{Department of Physics, University of California,
Davis, California 95616}

\date{\today}

\begin{abstract}
Replacing U for Ca in semiconducting CaB$_6$ at the few at.$\%$
level induces metallic behaviour and Kondo-type phenomena at low
temperatures, a rather unusual feature for U impurities in
metallic hosts. For Ca$_{0.992}$U$_{0.008}$B$_6$, the resistance
minimum occurs at $T$ = 17 K. The subsequent characteristic
logarithmic increase of the resistivity with decreasing
temperature merges into the expected $T^2$ dependence below 0.8 K.
Data of the low-temperature specific heat and the magnetization
are analyzed by employing a simple resonance-level model.
Analogous measurements on LaB$_6$ with a small amount of U
revealed no traces of Kondo behavior, above 0.4 K.
\end{abstract}
\maketitle

The Kondo phenomenon, a many-body effect affecting the conduction
electrons via their interaction with localized magnetic moments,
is notorious for d-transition metal impurities in simple metals
\cite{heeger}. Kondo behaviour is also often observed in metallic
compounds, in which a small number of cation sites is occupied by
rare-earth ions, usually trivalent Ce. A well known case for this
type of dilute Kondo systems is (La,Ce)B$_6$, with a Ce-content on
the few percent level. At low temperatures, all the typical Kondo
anomalies, as for example in the resistivity \cite{winzer} or in
the specific heat \cite{gruhl}, were observed. By adding the light
actinide element U instead of the lanthanide Ce into the same
metallic matrix does not lead to any Kondo anomalies, however
\cite{wiggi}. This reflects the notorious observation that
magnetic moments due to 5f electrons do not induce a Kondo effect
in a metallic host with a large number of itinerant charge
carriers. It is assumed that this is caused by a broadening of the
5f-states via hybridization with the conduction band states, and
hence the f-electrons loose their localized character. In this
work we demonstrate that the situation changes if U occupies a few
cation sites in the low-carrier density matrix CaB$_6$, with a
background density of conduction electrons of the order of
10$^{-4}$ per unit cell \cite{konrad}. Our data on the
low-temperature behaviour of the electrical resistivity, the
magnetic susceptibility and magnetization as well as the specific
heat clearly indicate that 5f electron moments may, under special
circumstances, induce the classical Kondo effect.

A single-crystalline sample of Ca$_{0.992}$U$_{0.008}$B$_6$ was
grown in a flux of Al, using the necessary high-purity starting
elements U, Ca and B. For measurements of the resistivity between
0.4 and 300 K, the sample was contacted at four small spots by a
silver-epoxy glue and a low-frequency ac-method was applied. The
specific heat was measured between 0.4 and 12 K, using a
relaxation-type method. Measurements of the susceptibility and the
magnetization, up to 5.5 T, were performed using a commercial
SQUID magnetometer. In order to avoid the influence of possible
magnetic impurities at the surface, the sample was etched several
times for a short duration in diluted nitric acid.

\begin{figure}
  \centering
  \includegraphics[width=\linewidth]{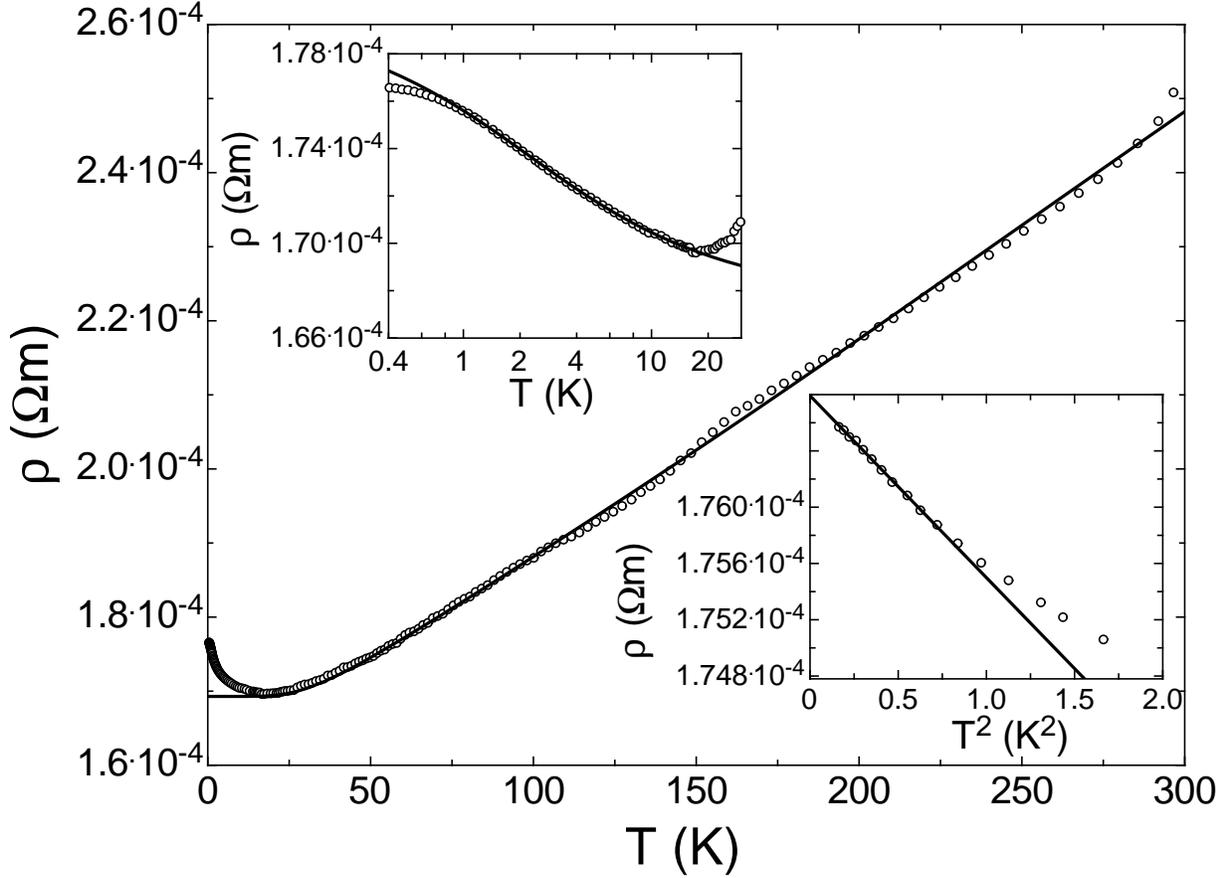}
  \caption{Temperature dependence of the electrical resistivity for Ca$_{0.992}$U$_{0.008}$B$_6$ in zero magnetic field. The
  solid line in the main frame represents the result of the calculation explained in the text.
  The left upper inset shows $\rho(T)$ for $T$ $<$ 30 K and the fit
  employing eq. \ref{hamannequation}. The right lower inset emphasizes the $T^2$ dependence of $\rho$
  for $T$ $<$ 0.8 K.}
 \label{resistivityplot}
\end{figure}

The main frame of Fig. \ref{resistivityplot} shows the temperature
dependence of the resistivity $\rho(T)$ for
Ca$_{0.992}$U$_{0.008}$B$_6$ in zero applied magnetic field and
for temperatures between 0.4 and 300 K. Below room temperature,
$\rho$ decreases almost linearly with decreasing $T$, reaches a
minimum at $T$ = $T_{min}$ $\approx$ 17 K and subsequently
increases again with decreasing temperature, tending to a finite
value at $T$ = 0 K.

At $T$ $>$ 30 K, $\rho(T)$ is best described by employing a model
that was recently introduced for handling Eu-based hexaborides
\cite{wiggermonnier,wigger}. The conduction electrons are assumed
to be scattered by disordered magnetic moments, lattice vibrations
and lattice defects. Considering the low concentration of magnetic
U ions and temperatures exceeding 30 K, our analysis establishes a
constant spin-disorder resistivity and a defect resistivity
$\rho_d$ = 1.68$\cdot$10$^{-4}$ $\Omega$m. The phonon spectrum of
the B$_6$-lattice is assumed to be that of a Debye solid with a
Debye temperature $\Theta_D$ = 1160 K \cite{mandrus}. The motion
of the Ca and U ions is taken into account by two independent
harmonic oscillators with Einstein-temperatures of
$\Theta_{E}^{Ca}$ = 373 K \cite{wigger} and $\Theta_{E}^{U}$ = 137
K, respectively. The solid line in Fig. \ref{resistivityplot}
represents the fit to $\rho(T)$ according to this calculation.

Below $T$ $\approx$ 17 K, the resistivity increases again with
decreasing $T$. Between 2.5 K and 10 K, this increase is very
close to logarithmic. In non-magnetic host materials with magnetic
impurities, such a $T$-dependence is usually attributed to the
formation of virtual bound states of the conduction electrons at
the magnetic sites. According to Hamann \cite{hamann}, the
magnetic impurity scattering leads to a resistivity

\begin{equation}\label{hamannequation}
  \rho_{imp} =
  \frac{\rho_0}{2}\left\{1-\frac{ln(T/T_K)}{[(ln(T/T_K))^2+\pi^2S(S+1)]^{1/2}}\right\}\ ,
\end{equation}

where $T_K$ is the Kondo temperature and $S$ is the spin of the
magnetic impurity. The upper inset in Fig. \ref{resistivityplot}
shows $\rho(T)$ for $T$ $<$ 30 K. The fit according to eq.
\ref{hamannequation}, displayed as the solid line in the inset,
yields $T_K$ = 2.0 $\pm$ 0.1 K and $S$ = 1/2, indicating a doublet
ground state of the crystal-field split U 5f electron multiplet.

In compounds, U adopts a tri- or tetravalent configuration. Under
the simplifying assumption that Hund's rule and Russel-Saunders
coupling are valid, free U$^{4+}$-ions carry a total angular
momentum of $J$ = 4. Lea, Leask and Wolf \cite{lea} established
the ground states for various values of $J$ in a variety of
different cubic crystal field environments. For $J$ = 4 the ground
state is either a singlet or a triplet. Free U$^{3+}$-ions adopt
an angular momentum $J$ = 9/2. For a value of the crystal field
parameter $x$ $<$ 0.4, the CEF split ground state is the doublet
state $\Gamma_6$. A spin value of $S$ = 1/2 obtained from fitting
eq. \ref{hamannequation} to our data suggests that U most likely
adopts the trivalent 5f$^3$ configuration and the ground state is
the $\Gamma_6$-doublet.

It is expected that at $T$ $\ll$ $T_K$, the Kondo system exhibits
a crossover to Fermi liquid behavior \cite{nagaoka}. For the
$n$-channel Kondo model with $n$ = 2$S$, the relevant contribution
to $\rho$ is \cite{yoshimori}

\begin{equation}\label{yoshimoriequation}
  \rho_{imp}(T) =
  \rho_{imp}(0)\left\{1-\left(\frac{\pi^2}{12T_L}\right)^2(4n+5)T^2\right\}
  \ \ .
\end{equation}

In our case, $n$ = 1 and the corresponding fit, shown in the lower
inset of Fig. \ref{resistivityplot}, results in $T_L$ $\sim$
$T_K\cdot\pi$ $\approx$ 6.0 K. This corresponds to $T_K$ $\approx$
1.9 K, in very good agreement with the value mentioned above.
Since $\rho_{imp}$ strongly depends on the magnitude of the
residual resistivity, this analysis is less reliable, however.

\begin{figure}
  \centering
  \includegraphics[width=\linewidth]{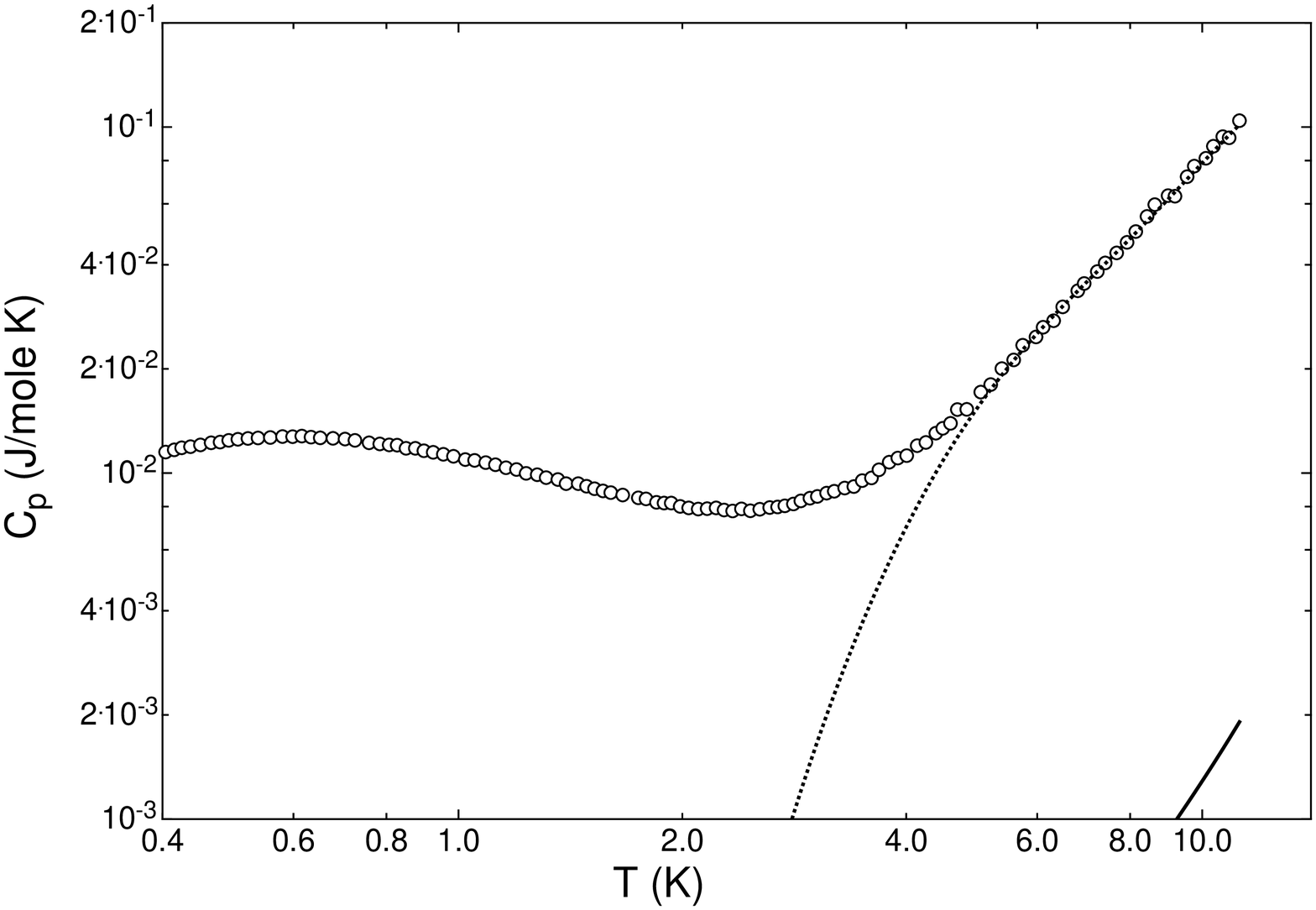}
  \caption{$C_p(T)$ in zero applied field for Ca$_{0.992}$U$_{0.008}$B$_6$ between 0.4 and
  12 K. The solid line represents the calculated phonon specific
  heat, the dotted line is the magnetic specific heat from the
  next higher magnetic levels.}
 \label{specheatplot}
\end{figure}

\begin{figure}
  \centering
  \includegraphics[width=\linewidth]{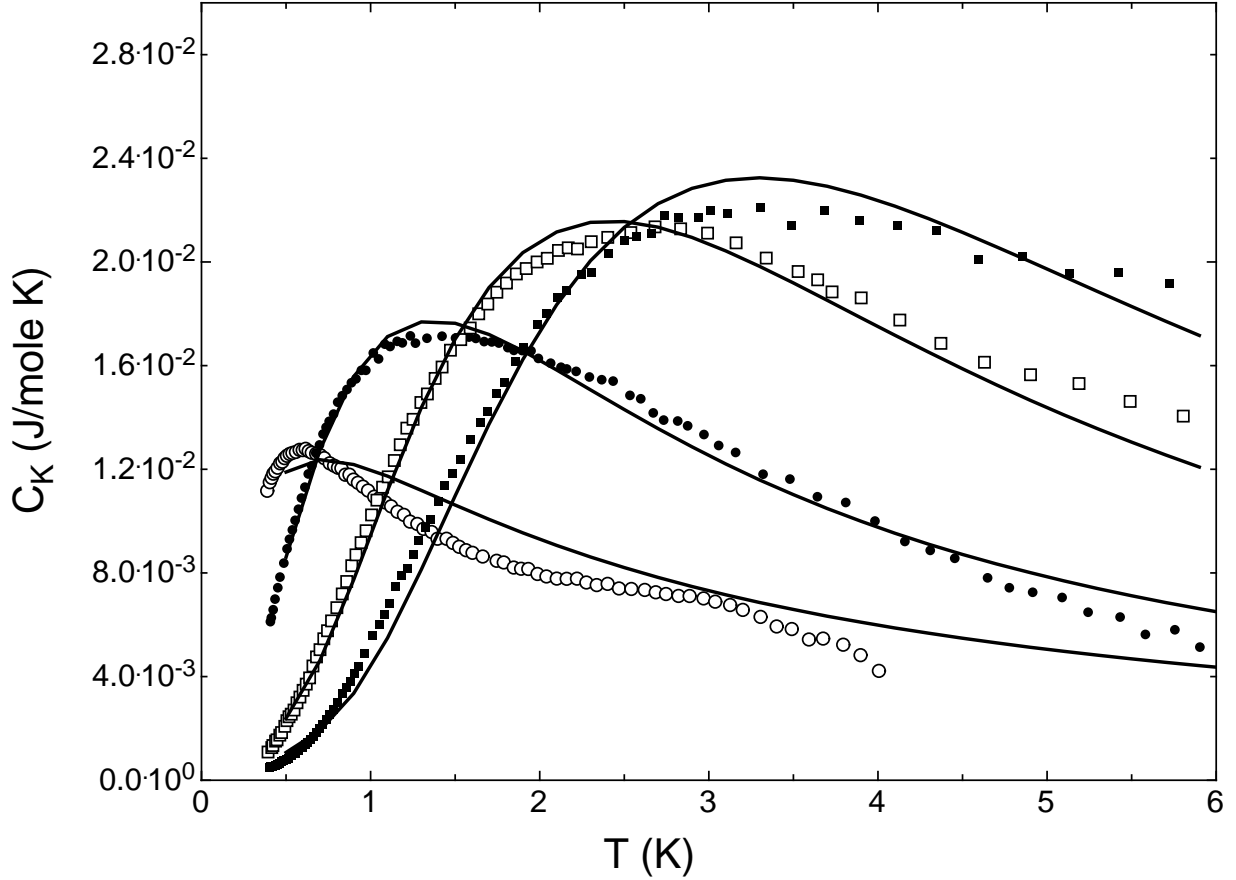}
  \caption{Kondo-induced electronic part of the specific heat $C_{K}(T,H)$ for
  Ca$_{0.992}$U$_{0.008}$B$_6$ between 0.4 and 6 K and at magnetic
  fields of $H$ = 0 kOe (empty circles), 25 kOe (filled circles), 50 kOe
  (empty squares) and 70 kOe (filled squares). The solid lines
  show the calculations according the resonance-level model
  \cite{schotte} described in the text.}
 \label{magspecheatplot}
\end{figure}

Further evidence for the Kondo behavior of U in CaB$_6$ is
obtained from measurements of the specific heat at low
temperatures. Fig. \ref{specheatplot} shows, on double-logarithmic
scales, the measured specific heat $C_p(T)$ in zero magnetic field
for the same specimen of Ca$_{0.992}$U$_{0.008}$B$_6$ between 0.4
and 12 K. The broad bell-shaped anomaly, observed at low $T$ with
a maximum at $T$ = 0.61 K, indicates the formation of a ground
state with a strongly enhanced electronic specific heat. The
background of the lattice specific heat was calculated by invoking
the same model parameters that were used in our calculation of
$\rho(T)$ described above. The result is represented by the solid
line in Fig. \ref{specheatplot}. For $T$ $>$ 6 K, we attribute the
excess specific heat to excitations to higher lying crystal field
levels of the 5f electrons. The corresponding fit, indicated by
the dotted line in Fig. \ref{specheatplot}, suggests that the next
higher levels are separated from the doublet ground state by
$\Delta/k_B$ = 25 K. It may be seen that at low temperatures,
$C_p(T)$ is dominated by the electronic contribution. Subtracting
the mentioned two background contributions from the experimental
$C_p(T)$ data results in $C_K(T)$

\begin{equation}
  C_{K}(T) = C_p(T) - C_{lattice}(T) - C_{CEF}(T) \ \ ,
\label{specheateq}
\end{equation}

which is plotted as open circles in Fig. \ref{magspecheatplot}.
This contribution is interpreted as being due to the Kondo-induced
enhanced density of states at the Fermi energy $D(E_F)$. Also
shown in Fig. \ref{magspecheatplot} are $C_K(T)$ data in different
external magnetic fields.

The maximum of the bell-shaped curve of $C_{K}(T)$ is, as
expected, shifted in magnitude and temperature upon application of
external magnetic fields. Our analysis of these data is based on
the resonance level (RL) model of Schotte and Schotte
\cite{schotte}, which assumes a Lorentzian shape of the electronic
density of states $D(\epsilon)$ centered at the Fermi level and
given by

\begin{equation}\label{kondodensity}
  D(\epsilon) = \Delta/\pi(\epsilon^2+\Delta^2) \ \ .
\end{equation}

We obtain consistent fits to the experimental curves of
$C_{K}(T,H)$ by postulating $S$ = 1/2, setting $\Delta/k_B$ = 1.6
K and inserting a $g$-factor of 1.7. They are shown as solid lines
in Fig. \ref{magspecheatplot}. The reduction of the $g$-factor is
equivalent to a reduced effective magnetic moment $\mu_{eff}$ of
the U ions, reflecting the magnetic screening of the local moments
by the conduction electrons \cite{schotte,schotte2}. Inserting the
value for $\Delta$ into eq. \ref{kondodensity} yields, for
$\epsilon$ = 0, the density of states $D(E_F)$. Considering that
the electronic specific heat parameter $\gamma$ is given by

\begin{equation}\label{gammanormal}
   \gamma = \frac{2}{3}\pi^2k_B^2D(E_F) \ \ ,
\end{equation}

we obtain $\gamma$ = 92 mJ/(K$^2\cdot$mole-U). For comparison, we
calculate $\gamma '$ for the case of an ordinary conduction band
with a quadratic dispersion relation and populated by one electron
per U-ion. The effective mass is $m^*$ = 0.28$\cdot m_0$, the same
as for itinerant charge carriers in CaB$_6$ \cite{gelderen}. With
these assumptions, we obtain $\gamma '$ = 0.19
mJ/(K$^2\cdot$mole-U), implying that the Kondo interaction leads
to an enhancement of the specific heat parameter by

\begin{equation}\label{gammaenhanced}
  \gamma_K / \gamma ' \approx 480 \ \ .
\end{equation}

\begin{figure}
  \centering
  \includegraphics[width=\linewidth]{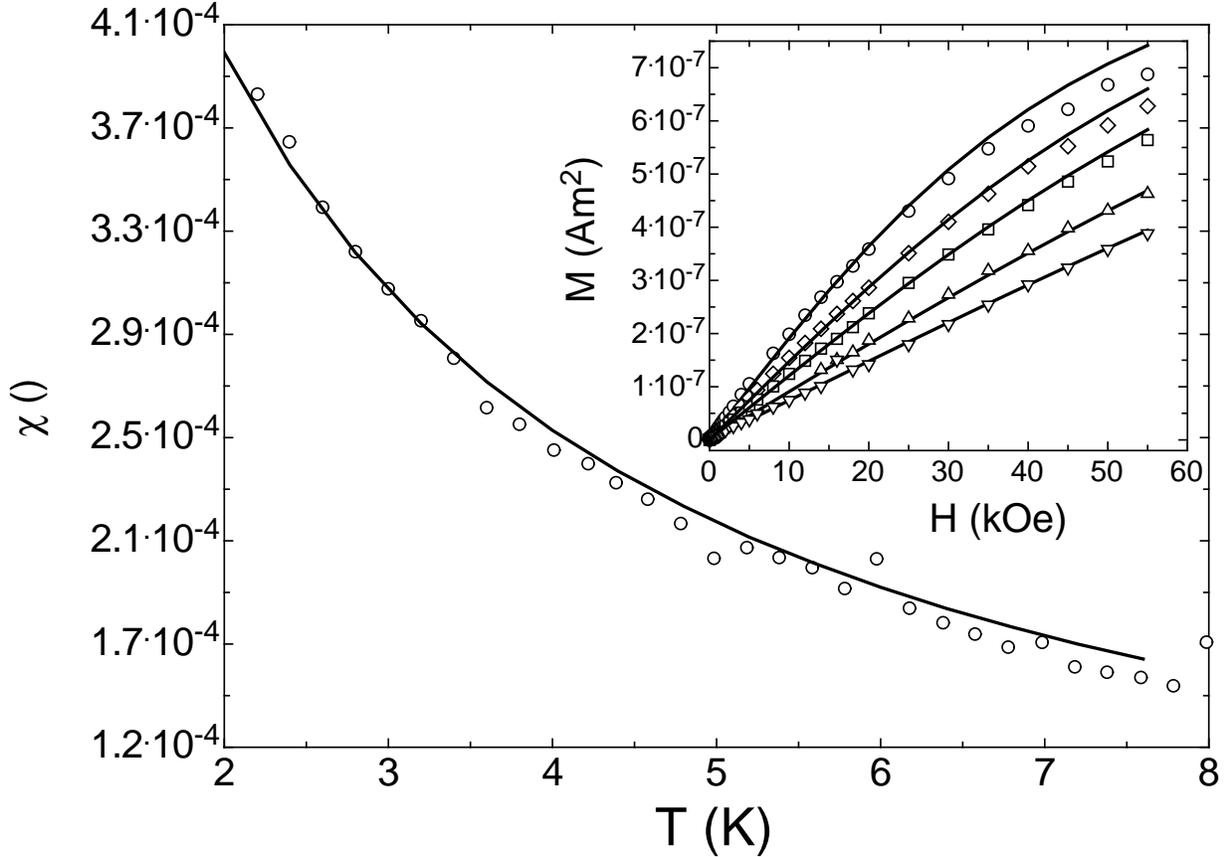}
  \caption{$\chi(T)$ at $H_{app}$ = 1 kOe for Ca$_{0.992}$U$_{0.008}$B$_6$. The inset shows $M(H)$
  for $T$ = 2 K (circles), 3 K (diamonds), 4 K (squares), 6 K
  (triangles up) and 8 K (triangles down). The solid lines
  describe the calculations according to the model of Schotte and
  Schotte described in the text.}
 \label{magnetizationplot}
\end{figure}

For $T$ $\gg$ $T_K$, the conduction electrons are incoherently
scattered by the actinide ions and the corresponding magnetic
moments may be regarded as isolated and only weakly coupled via
the RKKY interaction. Consequently, above 50 K, the susceptibility
$\chi(T)$, measured in an external magnetic field of 0.1 T and at
temperatures between 2 and 330 K, can be fitted by a Curie-Weiss
type law, with $J$ = 9/2, $g_J$ = 8/11 and a paramagnetic Curie
temperature of $\Theta_p$ $\approx$ - 5.9 K. The $T^{-1}$
dependence is riding above a constant paramagnetic background of
$\chi_0$ = 3.15$\cdot$10$^{-5}$. In our context, we concentrate on
the low-temperature features of $\chi(T)$ and $M(H)$. The
available data sets below 8 K are displayed in Fig.
\ref{magnetizationplot}. In order to be consistent with the
specific-heat analysis, the susceptibility $\chi(T)$, taken as
$M(H,T)/H$ at low magnetic fields $H$, was interpreted by again
employing the RL model of Schotte and Schotte \cite{schotte}.
Their general expression of $\chi(T)$ reduces, for $S$ = 1/2, to
the equation first presented by Rivier and Zuckermann
\cite{rivier}, who considered localized spin fluctuations in a
dilute non-magnetic alloy. The results of the calculation, using
the same parameters for the resonance width $\Delta$, the
$g$-factor and the spin $S$ as quoted above, is displayed as solid
lines in Fig. \ref{magnetizationplot} for both $\chi(T)$ in the
main frame, and $M(H,T)$ in the inset, respectively. It is
particularly rewarding that both magnetic and specific-heat data
can be well reproduced even quantitatively with this simple model.
Of course, all these considerations are only valid at low
temperatures, where the Kondo induced features dominate the
electronic properties.

Most earlier experimental attempts to demonstrate the onset of the
Kondo effect due to U impurities were either unsuccessful or did
not provide conclusive results. Previous work on the influence of
U replacing La in LaAl$_2$, e.g., claimed the first observation of
the Kondo effect in a dilute actinide alloy \cite{schlabitz}. We
believe that, in retrospect, the experimental evidence for that
claim is not convincing. We argue that also in that case, the
local moment character of the U ions was not well enough
developed. Some indications for a Kondo-type behaviour of the
resistivity \cite{haessler} and the low-temperature specific heat
\cite{haessler2} data were reported for compounds of the type
Th$_{1-x}$U$_x$Se and Th$_{1-x}$U$_x$S in the range 0.06 $<$ $x$
$<$ 0.26. The available data did not allow to unambiguously
distinguish between Kondo-behaviour \cite{haessler2} and a
mixed-valence situation \cite{haessler}. A logarithmic
$T$-dependence with a negative slope of $\rho(T)$ was also
reported for Th$_{0.86}$U$_{0.14}$Sb \cite{frick}. Again, the
situation seems ambiguous and a mixture of mixed-valency and Kondo
effect cannot be excluded \cite{nunes}. Our data, however,
indicate that the insertion of U at the low at. $\%$ level into a
matrix with a very low conduction electron concentration, such as
CaB$_6$, favors the stability of the local moment on the U site,
leading to a well developed classical Kondo effect at low
temperatures.

In conclusion, we have demonstrated, by means of resistivity,
specific heat and magnetization measurements, that a Kondo-type
behaviour by dilute impurities of U can be achieved by placing U
onto cation sites of the low-density itinerant carrier matrix
CaB$_6$. Replacing La by U in concentrations of the order of 1
$\%$ in LaB$_6$ does not lead to Kondo features at temperatures
above 0.4 K. In order to establish the reason for the formation of
localized 5f electrons on the U sites in CaB$_6$ but not in
LaB$_6$, additional information from, e.g., optical experiments
and theoretical considerations is required.

We appreciate stimulating discussions with R. Monnier. This work
has benefited from partial financial support of the Schweizerische
Nationalfonds zur F\"{o}rderung der wissenschaftlichen Forschung
and the US-NSF grant DMR-0433560.


\begin{thebibliography}{99}

\bibitem{heeger} A.J. Heeger, Solid State Physics Vol. 23, edited by F. Seitz, D.
Turnbull and H. Ehrenreich, (Academic, New York (1969)) p. 283.

\bibitem{winzer} K. Winzer, Sol. State Comm. {\bf 16}, 521 (1975).

\bibitem{gruhl} H. Gruhl and K. Winzer, Sol. State Comm. {\bf 57},
67 (1986).

\bibitem{wiggi} G.A. Wigger, unpublished.

\bibitem{konrad} K. Giann\`{o}, A.V. Sologubenko, H.R. Ott, A.D. Bianchi, and Z. Fisk,
 J. Phys.: Condens. Matter {\bf 14}, 1035 (2002).

\bibitem{wiggermonnier} G.A. Wigger, R. Monnier, H.R. Ott, D.P.
Young and Z. Fisk, Phys. Rev B {\bf 69}, 125118 (2004).

\bibitem{wigger} G.A. Wigger, E. Felder, H.R. Ott, A.D. Bianchi
and Z. Fisk, cond-mat/0404570.

\bibitem{mandrus} D. Mandrus, B. C. Sales and R. Jin, Phys. Rev. B {\bf 64},
12302 (2001).

\bibitem{hamann} D.R. Hamann, Phys. Rev. {\bf 158}, 570 (1967).

\bibitem{lea} K.R. Lea, M.J.M. Leask and W.P. Wolf, J. Phys. Chem.
Solids {\bf 23}, 1381 (1962).

\bibitem{nagaoka} Y. Nagaoka, Phys. Rev. {\bf 138}, 1112 (1965).

\bibitem{yoshimori} A. Yoshimori, J. Phys. C {\bf 15}, 5241 (1976)
$\&$ N. Mih\'{a}ly and A. Uawadowski, J. de Phys. Lett. {\bf 39},
L483 (1978).

\bibitem{schotte} K.D. Schotte and U. Schotte, Phys. Lett. {\bf 55A},
38 (1975).

\bibitem{schotte2} K.D. Schotte and U. Schotte, Phys. Rev. B {\bf 4},
2228 (1971).

\bibitem{gelderen} H.J. Tromp, P. van Gelderen, P.J. Kelly, G.
Brocks and P.A. Bobbert, Phys. Rev. Lett. {\bf 87}, 016401 (2000);
S. Massidda, A. Continenza, T.M. de Pascale and R. Monnier, Z.
Phys. B {\bf 102}, 83 (1997).

\bibitem{rivier} N. Rivier and M.J. Zuckermann, Phys. Rev. Lett
{\bf 21}, 904 (1968).

\bibitem{schlabitz} W. Schlabitz, F. Steglich, C.D. Bredl and W.
Franz, Physica B {\bf 102}, 321 (1980).

\bibitem{haessler} M. Haessler and C.-H. de Novion, J. Phys. C
{\bf 10}, 589 (1977);

\bibitem{haessler2} M. Haessler, M. Mortimer and C.-H. de
Novion, J. Phys. C: Solid State Phys. {\bf 16}, 1487 (1983).

\bibitem{frick} B. Frick, J. Schoenes, O. Vogt and J.W. Allen,
Solid State Commun. {\bf 42}, 331 (1982).

\bibitem{nunes} A.C. Nunes, J.W. Rasul, G.A. Gehring, J. Phys. C:
Solid State Phys. {\bf 19}, 1017 (1986).

\end{thebibliography}
\end{document}